\documentclass{emulateapj}

\usepackage{graphicx,natbib}
\usepackage{hyperref}

 \shorttitle{Apsidal motion and parameters of 21 SMC eclipsing binaries}
 \shortauthors{Zasche \& Wolf}

\begin{document}

   \title{Apsidal motion and absolute parameters of 21 early-type SMC \\ eccentric eclipsing binaries}

   \author{P. Zasche\altaffilmark{1}
          \and
          M. Wolf\altaffilmark{1} }

 \affil{
  \altaffilmark{1} Astronomical Institute, Charles University, Faculty of Mathematics and Physics, CZ-180~00, Praha 8, \\
             V~Hole\v{s}ovi\v{c}k\'ach 2, Czech Republic\\
              \email{zasche@sirrah.troja.mff.cuni.cz}  }


\begin{abstract}
 \noindent
We present the apsidal motion as well as the light curve analyses of 21 eccentric eclipsing
binaries located in the Small Magellanic Cloud. Most of these systems have never been studied
before, hence their orbital and physical properties as well as the apsidal motion parameters are
given here for the first time. All the systems are of early spectral type, having the orbital
periods up to 4 days. The apsidal motion periods were derived to be from 7.2 to 200 years
(OGLE-SMC-ECL-2194 having the shortest apsidal period among known main sequence systems). The
orbital eccentricities are usually rather mild (median of about 0.06), maximum eccentricity being
0.33. For the period analysis using the $O-C$ diagrams of eclipse timings, in total 951 minima were
derived from survey photometry as well as our new data. Moreover, 6 systems show some additional
variation in their $O-C$ diagrams, which should indicate the presence of hidden additional
components in them. According to our analysis these third-body variations have periods from 6.9 to
22 years.
 \end{abstract}

 \keywords{stars: binaries: eclipsing -- stars: fundamental parameters -- stars: early-type --
Magellanic Clouds}


\section{Introduction}

Eclipsing binaries still represent a most general method how to derive the basic stellar parameters
of the individual components such as radii, masses, or luminosities (see e.g.
\citealt{2012ocpd.conf...51S}). With these quantities, we would be able to improve and calibrate
the existing stellar evolution models (\citealt{2010A&ARv..18...67T}, or
\citealt{1997MNRAS.289..869P}). Moreover, the eclipsing binaries (hereafter EBs) can also serve as
suitable distance indicators, even for the close-by galaxies in the Local Group, see e.g.
(\citealt{2005MNRAS.357..304H}, or \citealt{2010A&A...509A..70V}).

A somewhat special group of EBs are those which have an eccentric orbit. For many such systems the
detected apsidal motion \citep{2010A&A...519A..57C} was analysed. Those systems were usually used
for studying the internal structure constants and also to test the general relativity
\citep{1993A&A...277..487C}. Moreover, for several apsidal motion systems also an additional third
body was detected, constituting even more dynamically interesting stellar system. For a compilation
of such systems, see a catalogue by \cite{2007MNRAS.379..370B}. Quite recently, a comprehensive
catalogue of 623 galactic eccentric EBs was published by \cite{2018ApJS..235...41K}, showing that
about 5\% of the systems with eccentric orbits have in general some third unseen components (only
based on eclipse timing variations, ETV). Additionally, also the processes like orbital
circularization \citep{2008EAS....29...67Z}, the distribution of period-eccentricity
(\citealt{2012ApJ...751....4K}, or \citealt{2005ApJ...620..970M}), or spin-orbit (mis)alignment can
be studied thanks to these objects.

Concerning the same objects outside our Galaxy the situation is slightly different. Due to much
lower luminosities, the data for extragalactic binaries only became available over the last two or
three decades. This is mainly due to the large photometric surveys like MACHO
(\citealt{1997ApJ...486..697A}, ranging the period 1992-2000), and OGLE
(\citealt{1992AcA....42..253U}, OGLE II ranging 1997-2000, OGLE III 2001-2009, and OGLE IV
2010-2014) for both Magellanic Clouds. These surveys harvested a huge portion of eclipsing
binaries, 6143 from the MACHO survey \citep{2007AJ....134.1963F}, and even 48605 from the OGLE
survey \citep{2016AcA....66..421P}. There are also many eccentric binaries with noticeable apsidal
motion among these systems.

This was a subject of several studies during the last couple of years. We have already published a
series of papers on apsidal motion in SMC and LMC: \cite{2013A&A...558A..51Z},
\cite{2014A&A...572A..71Z}, and \cite{2015AJ....150..183Z}. And besides that, the apsidal motion
was studied by other authors, most recently (and also most comprehensively) by
\cite{2016MNRAS.460..650H} presenting altogether 90 systems with apsidal motion in SMC, and
\cite{2015AJ....150....1H} studying 27 SMC systems.

Our present study is a natural continuation of such an effort. In Section \ref{methods} we present
the methods used for our analysis, while later in Section \ref{results} we introduce some of our
most interesting results, and finally in Section \ref{discussion} we briefly discuss the findings
and place them into a broader context.

\section{Methods of analysis}  \label{methods}

The huge majority of the targets presented in this study belongs to three groups. A first group,
consisting of five targets, contains objects which were already known to have eccentric orbits and
were already studied before. However, we collected many new observations of these systems during
the time span 2012-2018, hence we believe our analysis is better and more complete than the already
published ones. Second group comprises such stars, which were only by-chance discoveries in these
monitored fields of the stars from the first group and are also eccentric with apsidal motion. And
finally, eight systems were found scanning the OGLE III database systematically (star numbers
1-1000).

For all of these systems we tried to find all available data, i.e. photometry from the MACHO, OGLE,
and our new data from the 1.54-m Danish telescope located on La Silla Observatory in Chile
(hereafter DK154), equipped with the CCD camera, and $I$ and $R$ filters used (operated remotely
from the Czech Republic). Standard reduction procedure was applied for these new images, using the
bias and the flat fields to the CCD images. The comparison star was chosen to be close to the main
target and with a similar spectral type. A custom-made aperture-photometry reduction software {\sc
Aphot} developed by M.~Velen and P.~Pravec, was used for the data reduction. The correction for
differential extinction was not applied due to the close distance of the comparison star and the
variable one and the resulting negligible difference in air mass and their similar spectral types.
The other archival photometric data from OGLE and MACHO surveys were used in that way as they were
already published earlier on the websites or databases devoted to these surveys.

For analysing the light-curve (hereafter LC) of these binaries, we used the programme {\sc PHOEBE}
v0.32 \citep{2005ApJ...628..426P}, which is based on the Wilson-Devinney algorithm
(\citealt{1971ApJ...166..605W} and \citealt{1979ApJ...234.1054W}) and its later modifications.
Because of missing spectroscopic data for most of the targets, there are several limitations and
assumptions which need to be taken into account. At first, the mass ratios were kept fixed at
$q=1.0$. This approach is justified because all the systems are well-detached and the ellipsoidal
variations outside of their minima are generally negligible. For such systems their photometric
mass ratios can only hardly be derived, as quoted e.g. by \cite{2005ApSS.296..221T}. Only for these
systems where some spectroscopic data and the radial velocities were published, we naturally used
the mass ratio different from unity. We are aware of the fact that this assumption of $q=1.0$ is
rather odd for some of the systems where the light curve shapes show quite different eclipse depths
of both minima. Hence, we tried a test of using the method of deriving the photometric mass ratios
presented by \cite{2003MNRAS.342.1334G}, which uses the assumption of both components located on
the main sequence. For two systems from our sample which show significantly different eclipse
depths (namely OGLE-SMC-ECL-0752 and OGLE-SMC-ECL-1214), this method of deriving the mass ratio
from the luminosity ratio resulted in values of 0.80$\pm$0.08, and 0.61$\pm$0.06 respectively. As
one can see, both these numbers are away from unity, but still one can only speculate how these
values are close to reality when having no information about the radial velocities (and also
whether the main sequence assumption is appropriate or not). Having this alternative second LC fit,
we also performed the subsequent analysis (deriving the times of minima and carried out the period
analysis). The differences between this new solution and the original one with assumption $q=1.0$
are only marginal. The eccentricity and the apsidal motion period resulted in almost the same
values (differing by less than 2\%), well within their respective error bars.


The individual sources of information (MACHO, OGLE and our new Danish 1.54-m photometry) deal with
different photometric filters, namely $B$, $R$, and $I$ (the OGLE $V$ data were not used due to
much better OGLE $I$, both in quality and quantity). Different filters were fitted separately in
{\sc PHOEBE}, hence also different eclipse depths and luminosity ratios were derived. The values
given and discussed below in the Section \ref{results} are only these ones obtained for the $I$
filter because of the better quality and reliability. For the OGLE and MACHO data only the most
deviating data points (more than 3 sigma away) were omitted.

Here we summarize the individual steps of the analysis:
 \begin{itemize}

   \item At the beginning a preliminary light curve analysis was carried out (Initial
   OGLE period used, only rough estimation of the eccentricity, etc.).

   \item Initial LC analysis was used to derive preliminary minima, which were then
   analyzed to roughly estimate the apsidal motion (with the assumption $i=90^\circ$) and
   to judge whether to include the system into our sample or not (see Section \ref{discussion} below).

   \item Then, the eccentricity ($e$), argument of periastron ($\omega$) and apsidal motion
   rate ($\dot \omega$) resulting from the apsidal motion analysis were used for the
   light curve analysis.

   \item A value of inclination ($i$) from the LC analysis, was then used for the apsidal motion analysis.

   \item And finally, the resulting $e$, $\omega$, and $\dot \omega$ values from the apsidal motion
   analysis were kept fixed for the final LC analysis.
 \end{itemize}

A so-called AFP (Automatic Fitting Procedure) method was routinely used for deriving the times of
minima from the photometric surveys as well as our new data. It uses the whole light curve shape as
a template to calculate precise time of mid-eclipse. The method itself is presented in
\cite{2014A&A...572A..71Z}. All the minima used for the analysis are given in Table \ref{MINIMA}.

\begin{table}
 \caption{Heliocentric times of minima used for analysis.}
 \label{MINIMA}
 \scriptsize
 \centering
 \begin{tabular}{l l l c c l}
 \hline\hline
 Star              & HJD - 2400000 & Error  &   Filter & Reference\\
 \hline
 OGLE-SMC-ECL-0397 & 49177.26427 & 0.00266 & BR & MACHO \\ 
 OGLE-SMC-ECL-0397 & 49178.03816 & 0.00207 & BR & MACHO \\ 
 OGLE-SMC-ECL-0397 & 49651.22368 & 0.00204 & BR & MACHO \\ 
 OGLE-SMC-ECL-0397 & 49651.99717 & 0.00238 & BR & MACHO \\ 
 OGLE-SMC-ECL-0397 & 49948.40351 & 0.00336 & BR & MACHO \\ 
 OGLE-SMC-ECL-0397 & 49949.17003 & 0.00059 & BR & MACHO \\ 
 \dots \\
 \hline \hline
\end{tabular}
 \begin{list}{}{}
 \item Note: Table is published in its entirety in the electronic supplement of the journal.
 A portion is shown here for guidance regarding its form and content.
 \end{list}
\end{table}

\begin{table*}
\caption{Relevant information for the analysed systems.}  \label{InfoSystems}
 \tiny
  \centering \scalebox{0.99}{
\begin{tabular}{cccccccccccc}
   \hline\hline\noalign{\smallskip}
  \multicolumn{3}{c}{S\,y\,s\,t\,e\,m\,\,\,\, n\,a\,m\,e} &  RA  &    DEC    &$V_{\rm max}^{\,A}$& $(B-V)^{B}$&  $(B-V)_0$ & Sp.Type \\
  OGLE III       & OGLE II            & MACHO   &                &             & [mag]  & [mag]  & [mag]  &             \\
  \hline\noalign{\smallskip}
 \object{OGLE-SMC-ECL-0397} & SMC-SC2 11454 & 213.15277.84 &  00:39:33.91 & -73:18:55.8 & 17.279 & -0.015 & -0.294 &             \\ 
 \object{OGLE-SMC-ECL-0465} & SMC-SC2 33379 & 213.15389.115&  00:40:49.05 & -73:27:54.8 & 17.253 &  0.389 & -0.519 &             \\ 
 \object{OGLE-SMC-ECL-0639} &               &              &  00:43:16.20 & -73:41:51.9 & 17.824 &  0.051 & -0.163 &             \\ 
 \object{OGLE-SMC-ECL-0653} & SMC-SC3 92887 & 213.15565.189&  00:43:22.83 & -73:05:11.2 & 17.516 & -0.07  & -0.113 &             \\ 
 \object{OGLE-SMC-ECL-0752} & SMC-SC3 161183& 212.15625.117&  00:44:26.10 & -72:56:48.0 & 17.200 & -0.112 & -0.251 &             \\ 
 \object{OGLE-SMC-ECL-0787} & SMC-SC3 168989&              &  00:44:43.83 & -72:48:18.7 & 16.808 &  0.40  & -0.069 &             \\ 
 \object{OGLE-SMC-ECL-0874} & SMC-SC3 208420& 212.15678.74 &  00:45:26.89 & -73:10:05.1 & 16.328 &  0.10  & -0.244 &             \\ 
 \object{OGLE-SMC-ECL-0929} & SMC-SC4 47082 & 208.15685.160&  00:45:44.60 & -72:42:27.0 & 17.783 & -0.14  & -0.174 &             \\ 
 \object{OGLE-SMC-ECL-1214} & SMC-SC4 121461& 212.15792.439&  00:47:24.66 & -73:09:36.0 & 17.864 & -0.284 & -0.074 & B2$^C$      \\ 
 \object{OGLE-SMC-ECL-1370} & SMC-SC4 160094& 212.15847.154&  00:48:10.16 & -73:19:37.5 & 16.986 & -0.020 & -0.251 & B1$^C$      \\ 
 \object{OGLE-SMC-ECL-1393} & SMC-SC4 175333& 212.15850.358&  00:48:15.35 & -73:07:05.7 & 17.648 &  0.10  & -0.256 & B2$^C$      \\ 
 \object{OGLE-SMC-ECL-1558} & SMC-SC5 7078  &              &  00:49:02.31 & -73:27:48.2 & 14.108 & -0.12  & -0.269 & B0-5(II)$^D$\\ 
 \object{OGLE-SMC-ECL-1634} & SMC-SC5 123390&212.15908.2537&  00:49:22.65 & -73:03:43.2 & 16.023 &  0.05  & -0.278 & B1$^C$      \\ 
 \object{OGLE-SMC-ECL-1874} & SMC-SC5 185408& 212.15962.211&  00:50:24.53 & -73:14:56.4 & 17.405 & -0.031 & -0.266 & B2$^C$      \\ 
 \object{OGLE-SMC-ECL-1985} & SMC-SC5 225507& 208.16025.300&  00:50:49.11 & -72:50:21.7 & 17.629 & -0.034 & -0.207 &             \\ 
 \object{OGLE-SMC-ECL-2112} & SMC-SC5 266084&              &  00:51:18.84 & -73:14:01.8 & 17.180 &  0.09  & -0.195 &             \\ 
 \object{OGLE-SMC-ECL-2152} & SMC-SC5 265970& 212.16076.59 &  00:51:28.10 & -73:15:18.0 & 16.121 &  0.03  & -0.230 & B1$^C$      \\ 
 \object{OGLE-SMC-ECL-2194} & SMC-SC5 266131& 212.16077.197&  00:51:35.80 & -73:12:45.2 & 17.071 &  0.056 & -0.280 & B1$^C$      \\ 
 \object{OGLE-SMC-ECL-2385} & SMC-SC6 67902 & 208.16084.193&  00:52:12.12 & -72:44:53.6 & 17.126 &  0.445 & -0.611 &             \\ 
 \object{OGLE-SMC-ECL-2460} & SMC-SC6 100626& 212.16133.476&  00:52:29.95 & -73:16:51.6 & 18.526 &  0.103 & -0.150 &             \\ 
 \object{OGLE-SMC-ECL-4923} & SMC-SC10 41690&              &  01:04:39.47 & -72:49:49.8 & 15.952 & -0.20  & -0.283 &             \\ 
 \noalign{\smallskip}\hline
\end{tabular}}\\
\scriptsize Note: [A] - Out-of-eclipse $V$ magnitude based on OGLE database, see
\cite{2013AcA....63..323P}; [B] - photometric index by \cite{2002AJ....123..855Z} or
\cite{2002ApJS..141...81M}; [C] - spectral type from \cite{2010A&A...520A..74N}, and [D] - from
\cite{2004MNRAS.353..601E}.
\end{table*}

\begin{table*}
 \caption{The parameters of the light curve fits and the apsidal motion.}
 \label{LCOCparam}
 \tiny
 \centering \scalebox{0.68}{
 \begin{tabular}{l c r c c c c c c c r c c c c c c c}
 \hline\hline
  System           &    $i$       & \multicolumn{1}{c}{$T_1$} & \multicolumn{1}{c}{$T_2$} &  $L_1$  &  $L_2$  & $L_3$  & $R_1/a$ & $R_2/a$ & $HJD_0     $ & \multicolumn{1}{c}{$P$ [d]} &  $e$   & $\omega$ [deg] & $U$ [yr] \\
                   &    [deg]     & \multicolumn{1}{c}{[K]}   & \multicolumn{1}{c}{ [K]}  &  [\%]   &   [\%]  &  [\% ] &         &         & $[2400000+]$ &            &        &                &          \\
 \hline
 OGLE-SMC-ECL-0397 & 90.96 (0.98) & 29000 & 28238 (370) & 51.5 (1.9) & 48.5 (1.7) & 0.0       &  0.262 (2) & 0.259 (2) & 55503.3151 (14) & 1.5239810 (12) & 0.016 (4) &  24.3 (0.8) & 17.4 (0.9) \\ 
 OGLE-SMC-ECL-0465 & 85.09 (1.22) & 15000 & 15743 (305) & 40.2 (1.4) & 57.0 (3.2) & 2.8 (3.0) &  0.256 (5) & 0.296 (4) & 55610.2068 (9)  & 1.7208977 (8)  & 0.009 (2) & 335.5 (0.9) & 15.4 (0.7) \\ 
 OGLE-SMC-ECL-0639 & 82.22 (0.75) & 16000 & 17121 (1216)& 42.0 (3.2) & 57.8 (3.8) & 0.2 (0.7) &  0.193 (7) & 0.217 (3) & 53501.2182 (68) & 2.2176130 (63) & 0.140 (19)&  95.8 (2.6) & 77.5 (16.9)\\ 
 OGLE-SMC-ECL-0653 & 86.58 (0.30) & 13000 & 11973 (349) & 60.3 (1.0) & 37.8 (1.5) & 1.8 (1.0) &  0.154 (3) & 0.133 (2) & 53502.2863 (209)& 3.8022773 (109)& 0.333 (87)& 216.8 (18.5)&170.1 (43.7)\\ 
 OGLE-SMC-ECL-0752 & 79.59 (0.63) & 24000 & 21353 (439) & 56.1 (2.3) & 40.9 (2.0) & 3.0 (1.8) &  0.187 (4) & 0.172 (2) & 55500.8284 (52) & 2.1130468 (77) & 0.114 (24)&  31.4 (4.9) & 78.0 (7.3) \\ 
 OGLE-SMC-ECL-0787 & 82.38 (1.06) & 10500 & 10590 (548) &  6.0 (0.8) & 20.9 (3.1) &73.1 (4.8) &  0.116 (2) & 0.235 (4) & 55402.2604 (86) & 2.5592298 (101)& 0.217 (55)&  48.8 (6.0) & 71.6 (8.2) \\ 
 OGLE-SMC-ECL-0874 & 80.16 (0.62) & 25000 & 20493 (498) & 29.4 (1.1) & 13.1 (0.3) &57.5 (2.3) &  0.219 (3) & 0.177 (4) & 55601.0605 (60) & 1.8846380 (52) & 0.109 (15)& 294.2 (3.2) & 56.7 (4.0) \\ 
 OGLE-SMC-ECL-0929 & 79.43 (0.57) & 17000 & 17630 (420) & 48.3 (1.0) & 51.7 (1.8) & 0.0       &  0.229 (5) & 0.231 (5) & 52003.2937 (23) & 2.1273835 (22) & 0.087 (12)& 104.1 (2.8) & 54.2 (3.2) \\ 
 OGLE-SMC-ECL-1214 & 82.44 (0.61) & 21000 & 20757 (629) & 68.5 (1.4) & 31.5 (0.9) & 0.0       &  0.188 (4) & 0.131 (3) & 52217.6935 (42) & 1.9467130 (12) & 0.148 (17)&  29.9 (3.1) & 63.4 (4.9) \\ 
 OGLE-SMC-ECL-1370 & 76.53 (0.22) & 25000 & 22437 (431) & 57.2 (3.2) & 36.1 (3.8) & 6.7 (2.9) &  0.217 (5) & 0.191 (3) & 53689.5886 (11) & 1.6996138 (13) & 0.086 (7) & 134.2 (2.3) & 29.1 (1.5) \\ 
 OGLE-SMC-ECL-1393 & 81.09 (0.90) & 21000 & 18211 (510) & 68.0 (2.3) & 32.0 (2.0) & 0.0       &  0.257 (4) & 0.190 (7) & 52473.2811 (6)  & 1.2511244 (4)  & 0.017 (3) & 305.1 (1.1) & 16.9 (0.9) \\ 
 OGLE-SMC-ECL-1558 & 71.06 (0.31) & 26400 & 22643 (335) & 40.9 (3.4) & 27.0 (1.5) &32.1 (2.8) &  0.223 (6) & 0.209 (7) & 52120.9320 (42) & 2.4472722 (45) & 0.035 (22)& 166.4 (5.2) & 46.4 (7.4) \\ 
 OGLE-SMC-ECL-1634 & 77.90 (1.15) & 28000 & 22920 (329) & 77.1 (0.9) & 12.6 (1.4) &10.3 (3.7) &  0.248 (3) & 0.129 (6) & 52132.9208 (22) & 2.1728736 (18) & 0.034 (25)&  91.8 (4.7) & 37.8 (2.3) \\ 
 OGLE-SMC-ECL-1874 & 75.98 (0.56) & 21000 & 19545 (454) & 60.1 (0.7) & 39.9 (0.5) & 0.0       &  0.253 (2) & 0.214 (3) & 55500.5525 (16) & 1.4549890 (9)  & 0.027 (5) & 187.7 (1.9) & 17.1 (0.9) \\ 
 OGLE-SMC-ECL-1985 & 75.04 (0.73) & 20500 & 21089 (688) & 60.7 (0.8) & 39.3 (0.9) & 0.0       &  0.244 (3) & 0.195 (3) & 51175.9685 (30) & 1.5109147 (16) & 0.048 (12)&   2.7 (8.6) & 29.2 (5.6) \\ 
 OGLE-SMC-ECL-2112 & 82.28 (1.89) & 18000 & 16971 (539) & 32.3 (7.0) & 26.5 (3.8) &41.2 (7.5) &  0.194 (9) & 0.182 (9) & 52500.5798 (59) & 2.1319415 (92) & 0.059 (58)&  92.5 (11.7)&190.2 (28.8)\\ 
 OGLE-SMC-ECL-2152 & 90.81 (1.01) & 23000 & 22413 (231) & 69.9 (2.6) & 21.1 (1.2) & 9.0 (4.3) &  0.265 (2) & 0.149 (4) & 53503.1812 (102)& 3.4956503 (112)& 0.061 (17)&   9.3 (5.3) & 72.6 (12.9)\\ 
 OGLE-SMC-ECL-2194 & 83.61 (0.65) & 26500 & 25280 (542) & 59.8 (1.9) & 40.2 (1.7) & 0.0       &  0.285 (6) & 0.243 (5) & 55001.0438 (12) & 1.3029425 (8)  & 0.042 (5) & 126.3 (1.0) &  7.2 (0.3) \\ 
 OGLE-SMC-ECL-2385 & 92.02 (0.78) & 20000 & 20010 (341) & 44.9 (2.8) & 50.8 (2.4) & 4.3 (2.6) &  0.198 (3) & 0.209 (7) & 51173.6235 (14) & 1.7186416 (14) & 0.058 (10)& 260.3 (1.2) & 35.4 (2.1) \\ 
 OGLE-SMC-ECL-2460 & 81.66 (1.04) & 15000 & 14825 (553) & 54.8 (3.0) & 45.2 (4.1) & 0.0       &  0.218 (3) & 0.201 (9) & 53000.3362 (76) & 2.0724939 (80) & 0.121 (22)&  43.1 (3.9) & 95.7 (9.8) \\ 
 OGLE-SMC-ECL-4923 & 82.36 (0.51) & 26000 & 31350 (751) & 43.4 (2.6) & 56.6 (2.4) & 0.0       &  0.138 (6) & 0.141 (4) & 50716.9851 (128)& 3.3114695 (104)& 0.312 (88)&  24.2 (10.0)& 53.0 (12.5)\\ 
 \hline
 \end{tabular}}
\end{table*}

\begin{table*}
\caption{The parameters of the third-body orbits for the individual systems.} \label{LITEparam}
 \scriptsize
\begin{tabular}{ccccccccc}
\hline\hline\noalign{\smallskip}
  System           &   $A$       & $\omega_3$   & $P_3$      & $T_0$ [HJD] &    $e_3$    & $f(m_3)$   & $P_3^2/P$ & $P^2/P_3^{5/3}$   \\                           
                   &  [days]     &   [deg]      & [yr]       & (2400000+)  &             & $[M_\odot]$& [yr]      & ($ \times  10^{-5}$) \\
 \noalign{\smallskip}\hline\noalign{\smallskip}
 OGLE-SMC-ECL-0874 & 0.0497 (68) & 342.0 (16.3) & 21.4 (0.2) & 69450 (225) & 0.727 (28)  & 3.71 (56)  &  88453    & 0.12              \\ 
 OGLE-SMC-ECL-1558 & 0.0250 (49) & 334.6 (21.3) & 14.3 (0.9) & 61228 (407) & 0.652 (104) & 0.75 (8)   &  30720    & 0.38              \\ 
 OGLE-SMC-ECL-1634 & 0.0110 (23) & 207.1 (4.3)  &  6.9 (0.1) & 58164 (112) & 0.473 (52)  & 0.19 (2)   &  7937     & 1.01              \\ 
 OGLE-SMC-ECL-2112 & 0.0295 (30) & 230.2 (24.0) & 11.9 (0.7) & 58979 (314) & 0.406 (97)  & 1.06 (14)  & 24093     & 0.39              \\ 
 OGLE-SMC-ECL-2152 & 0.0057 (14) & 254.6 (54.8) &  7.5 (2.4) & 56964 (661) & 0.005 (5)   & 0.017 (2)  &  5888     & 2.28              \\ 
 OGLE-SMC-ECL-2385 & 0.0020 (10) & 229.8 (48.2) &  9.2 (1.5) & 57975 (439) & 0.184 (101) & 0.0005 (1) & 17844     & 0.39              \\ 
  \hline
 \noalign{\smallskip}\hline
\end{tabular}
\end{table*}

\begin{figure*}
  \centering
  \includegraphics[width=0.8\textwidth]{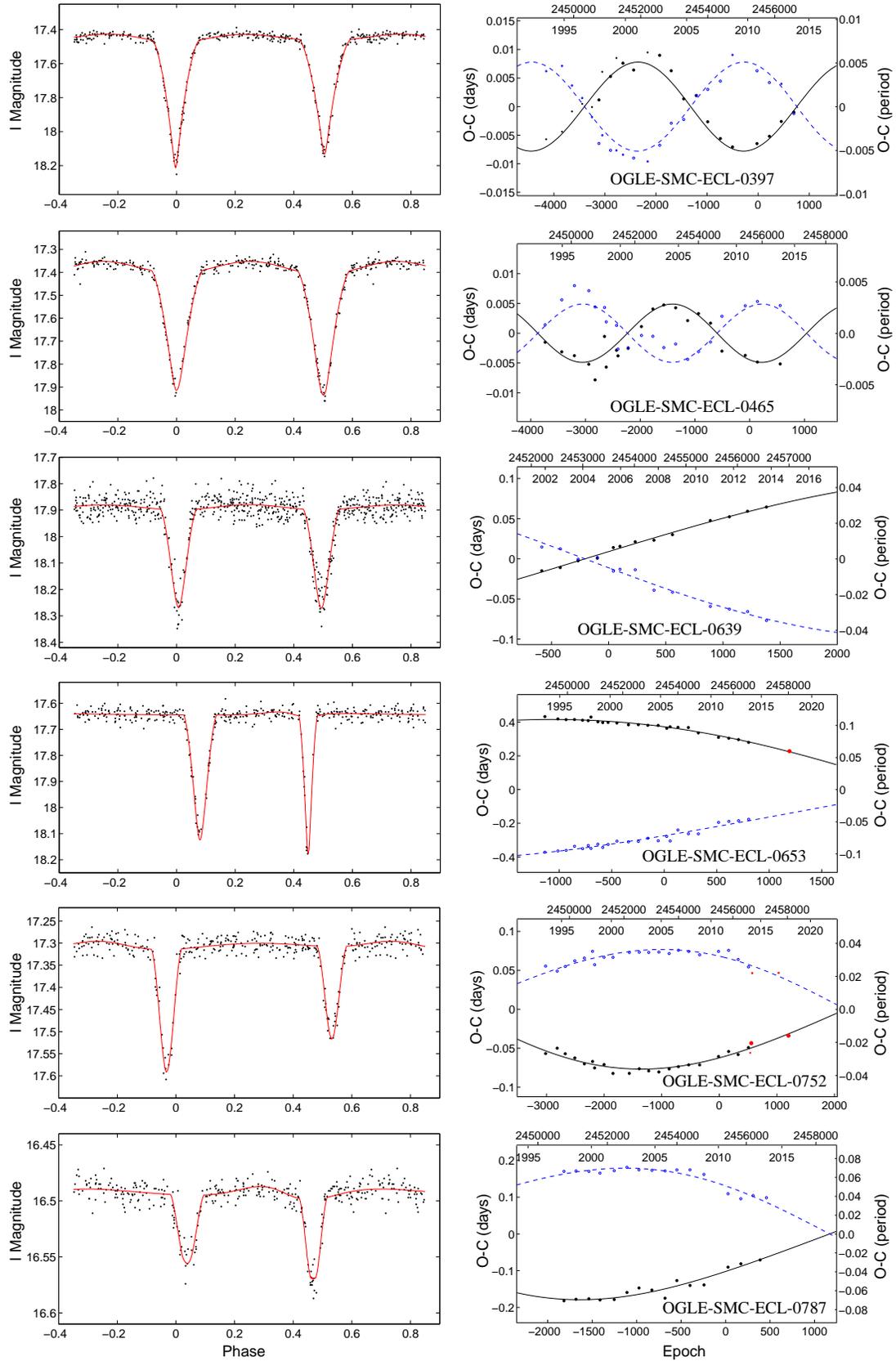}
  \caption{Plot of the light curves and $O-C$ diagrams of the analysed systems. For the $O-C$ diagrams
  the full dots stand for the primary minima (as well as the solid line), while the open circles
  represent the secondary minima (and the dashed curve). Red symbols stand for our new minima derived
  from photometry taken with Danish 1.54-m telescope in La Silla.}
  \label{FigLCOC1}
\end{figure*}

\begin{figure*}
  \centering
  \includegraphics[width=0.85\textwidth]{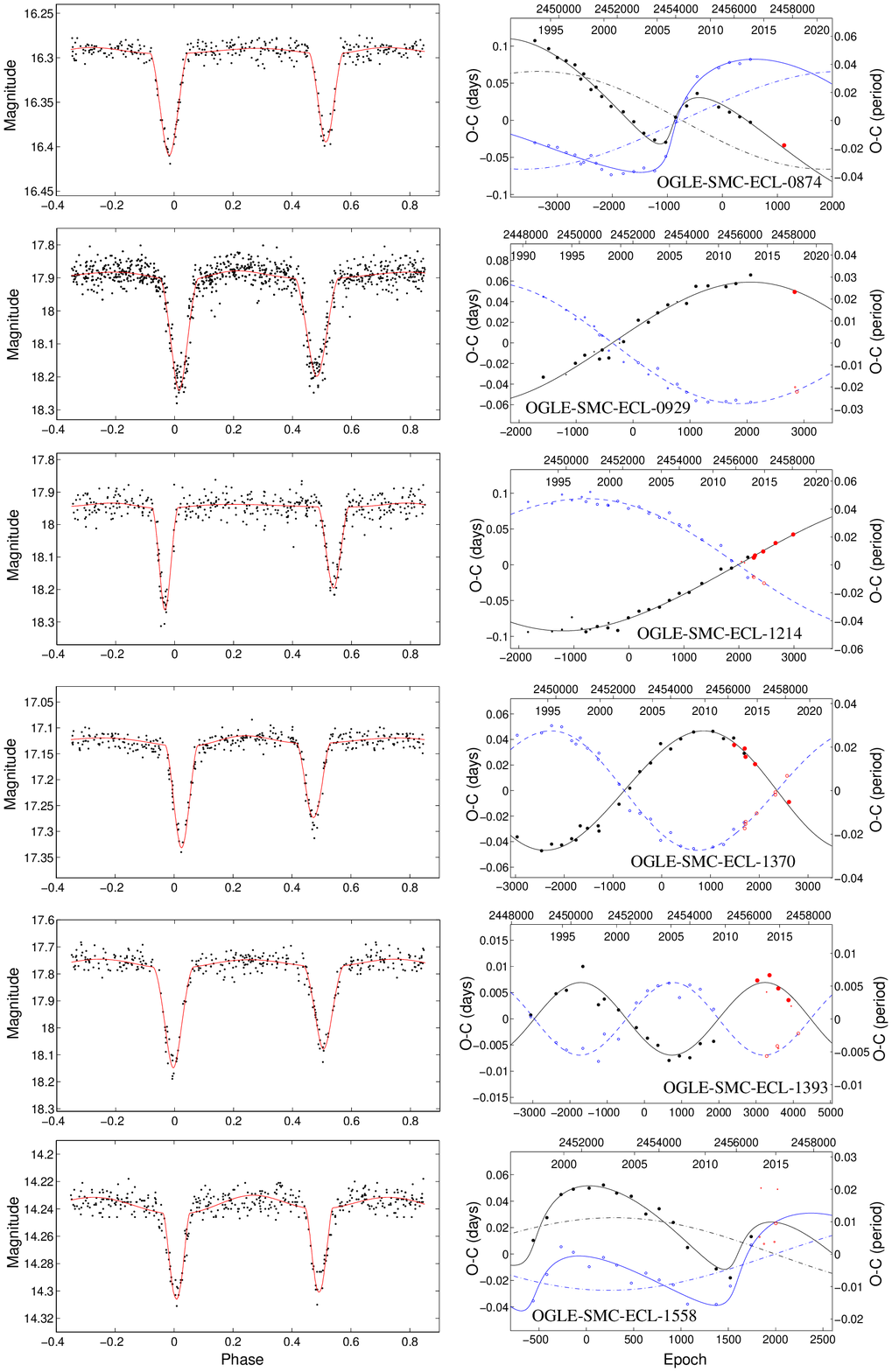}
  \caption{Plot of the light curves and $O-C$ diagrams, continuation.}
  \label{FigLCOC2}
\end{figure*}

\begin{figure*}
  \centering
  \includegraphics[width=0.85\textwidth]{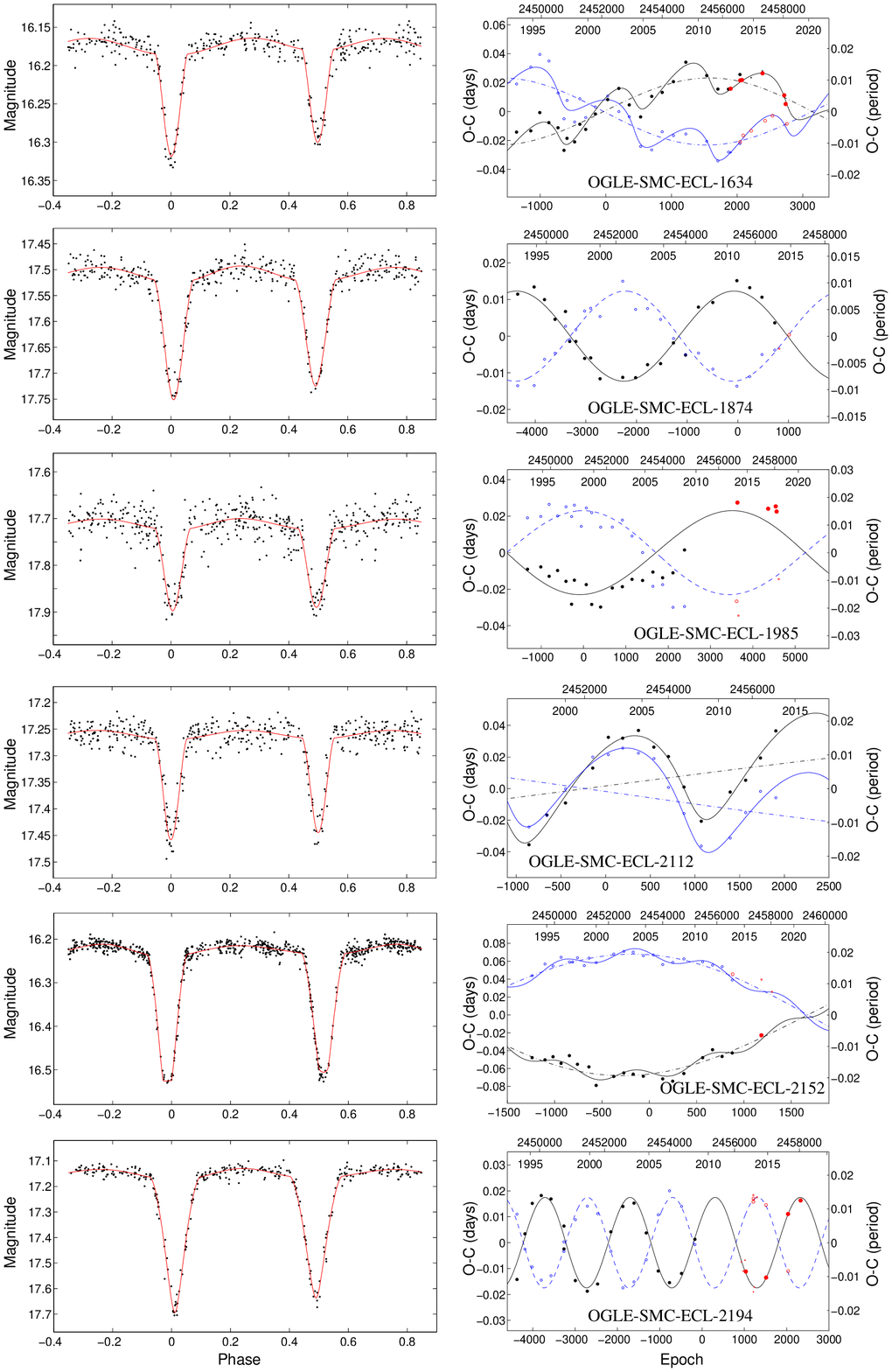}
  \caption{Plot of the light curves and $O-C$ diagrams, continuation.}
  \label{FigLCOC3}
\end{figure*}

\begin{figure*}
  \centering
  \includegraphics[width=0.85\textwidth]{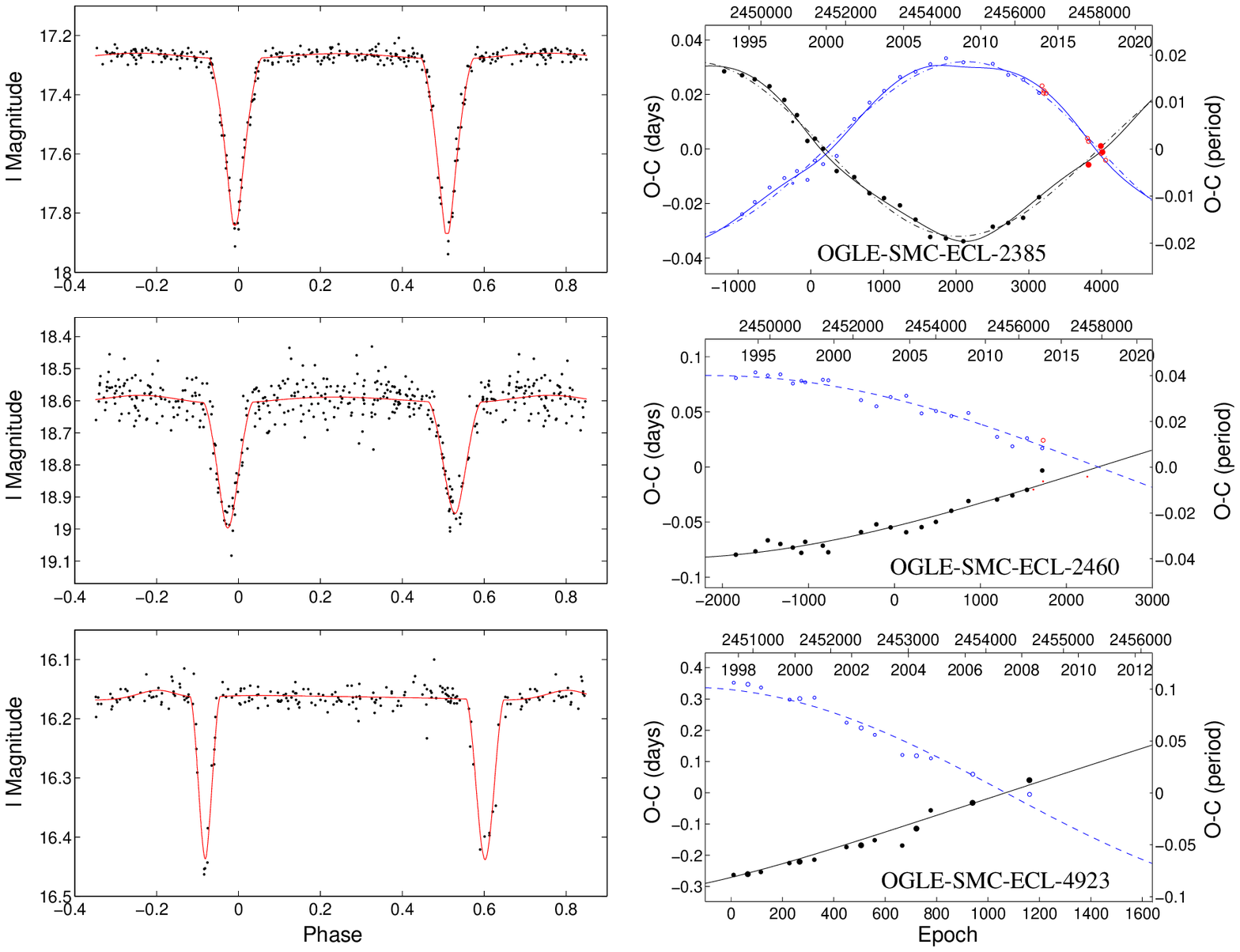}
  \caption{Plot of the light curves and $O-C$ diagrams, continuation.}
  \label{FigLCOC4}
\end{figure*}

\begin{figure*}
  \centering
  \includegraphics[width=0.85\textwidth]{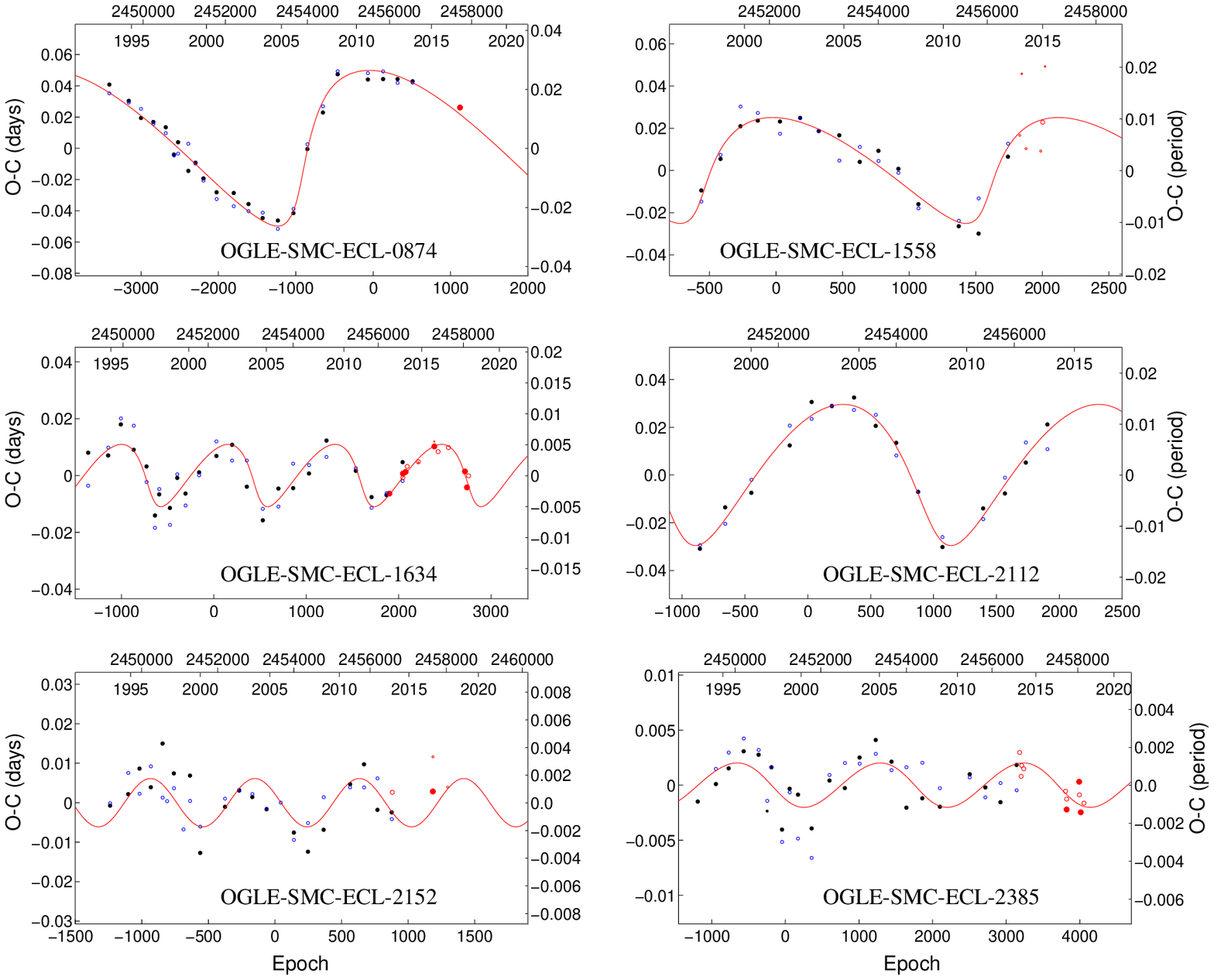}
  \caption{Plot of the $O-C$ diagrams with the light-time effect fits (after subtraction of the apsidal motion).}
  \label{FigOClite}
\end{figure*}

\section{The results}  \label{results}

As was already mentioned earlier in Section \ref{methods}, a whole portion of systems is rather
heterogeneous, hence it comprises also the stars already studied (which are rather brighter), as
well as those ones never studied before, which are slightly fainter. In Table \ref{InfoSystems} we
can see some basic information about the particular systems, where one can find besides the
alternative designations of the particular star also its coordinates for precise identification.
Besides that, the photometric indices as found in the various databases and surveys are given
together with the dereddened value of $(B-V)_0$. These were calculated from the individual $(B-V)$,
and $(U-B)$ indices following the method published in \cite{1958LowOB...4...37J}. However, in some
cases this method yielded in rather unreliable results (this is the case for e.g.
OGLE-SMC-ECL-0465, and OGLE-SMC-ECL-2385), hence even this spectral estimation cannot be used for
deriving the primary temperature $T_1$. In most cases where some spectral classification was
published earlier (by \citealt{2010A&A...520A..74N}, and \citealt{2004MNRAS.353..601E}) our
computed dereddened values of $(B-V)_0$ resulted in quite reasonable values. The individual
spectral types and their temperatures were assigned from the $(B-V)_0$ values according to tables
by \cite{2013ApJS..208....9P}.

For those systems where some spectroscopy was published the situation was a bit different. For
several systems their radial velocity curves exist, and the resulting mass ratio $q$ as published
by \cite{2010A&A...520A..74N} was used for our modelling. Moreover, for one system (namely
OGLE-SMC-ECL-1558) the spectral type classification together with the primary temperature value was
also published \citep{2004MNRAS.353..601E}, hence also this information was used. All the results
of our fitting are given in Table \ref{LCOCparam} and Figures \ref{FigLCOC1} to \ref{FigLCOC4}. In
these light curve graphs we plotted only the data from shorter time interval when the change of
$\omega$ is only small (for better clarity not to plot the blurred phase light curve over the
longer time interval due to precession of omega angle). In the Table \ref{LCOCparam} there are
given the inclination $i$, fixed value of primary temperature $T_1$, computed temperature $T_2$,
luminosity ratios in $I$ filter for all components, and relative radii of both stars -- all of
these as resulted from the {\sc PHOEBE} fitting. Moreover, also the parameters from the apsidal
motion analysis are given there, namely: linear ephemerides $HJD_0$, and $P$ together with the
eccentricity of the orbit $e$, argument of periastron $\omega_0$ at a reference time of $HJD_0$,
and also the apsidal period $U$.

As one can see, for a significant number of systems also an additional variation besides the
apsidal motion was detected. We attributed this variation to some hypothetical additional third
body in the system causing so-called light-time effect due to its orbital motion around a common
barycenter. This method was described elsewhere (e.g. \citealt{Irwin1959}, or \citealt{Mayer1990}),
and it is being almost routinely used nowadays for various surveys and satellite data, see e.g.
\cite{2016MNRAS.455.4136B}. For these systems, where an additional body was detected, their $O-C$
diagrams after subtraction of the apsidal motion were plotted with the light-time effect fits, see
Fig. \ref{FigOClite}. The parameters of these potential third-body orbits are given in Table
\ref{LITEparam}, where the amplitude of variation is denoted as $A$, while $\omega_3$ stands for
the argument of periastron of the third orbit, $T_0$ time of the periastron passage, $P_3$ its
period, and $e_3$ its eccentricity, respectively. Mass function is $f(m_3)$, and the last two
columns present the ratio of squared outer and inner periods, and the fraction $P^2/P_3^{5/3}$.
Classical geometrical light-time effect has its amplitude proportional to $\sim P_3^{2/3} \cdot
f(m_3)^{1/3}$, however, the dynamical perturbation is much more dominant in tight triples and its
amplitude is proportional to ${P^2/P_3} \cdot {m_3/m_{1+2+3}}$, see e.g.
\citealt{2016MNRAS.455.4136B} for details. On the other hand, the slow precession of the orbits can
possibly be detected only when the nodal period \citep{1975A&A....42..229S} is adequately short,
but its period is proportional to $P_3^2/P$. And as one can see, this ratio is too large for all of
our studied systems. And finally, also the ratio of amplitudes for both contributions
${A_{dyn}}/{A_{LITE}} \sim P^2/P_3^{5/3}$,  is adequately small for all of them.

Concerning the systems already studied in \cite{2016MNRAS.460..650H}, our presented fits seem to be
better mainly due to the fact that we deal with larger data set, spanning longer period. Hence, our
results should be more robust than the ones already published. The new data obtained with the
Danish 1.54-m telescope were secured during the particular observing runs and present definitely
more suitable photometry than the sparse OGLE photometric data (having typically only one data
point per night).

\begin{figure}[t]
 \vskip 2mm
  \centering
  \includegraphics[width=0.45\textwidth]{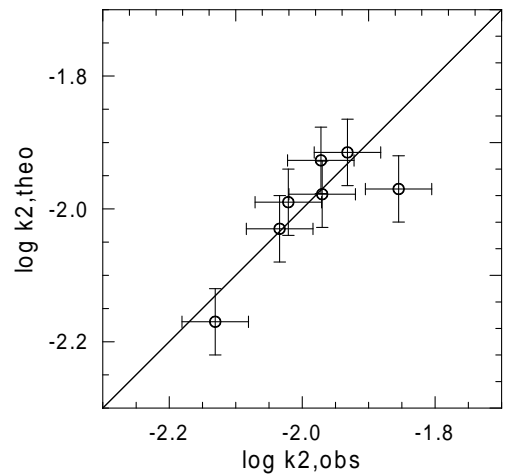}
  \caption{Internal structure constants, the observed ones compared with the theoretical ones.}
  \label{FigK2}
\end{figure}

We also computed the internal structure constants from the inferred apsidal motion. We calculated
these values for seven systems from our sample which were also studied spectroscopically (by
\citealt{2010A&A...520A..74N}), hence their masses are known precisely. The observed values were
compared with the theoretical ones published by \cite{2005A&A...440..647C} as one can see in Table
\ref{tabK2} and also Fig. \ref{FigK2}. For the derivation of the theoretical values we used the
same LMC metallicity as used by \cite{2010A&A...520A..74N}, as well as the stellar ages derived
from their analysis. The relativistic contribution to the total apsidal motion rate is also given
in Table \ref{tabK2}, and is relatively small for all of the systems.

 \section{Discussion and conclusions}  \label{discussion}

We have derived the preliminary apsidal motion and light curve parameters for 21 detached eclipsing
binaries. Among these systems, several ones have already been studied before. However, we harvested
also from other data sources, namely the spectroscopy (in contrast with the previous study by
\citealt{2016MNRAS.460..650H}), the long-term photometry for the better derivation of the apsidal
motion (in contrast with the previous spectroscopic study by \citealt{2010A&A...520A..74N}), as
well as our new precise photometry from the Danish 1.54-meter telescope.

The derived apsidal motion periods resulted in quite reasonable values of several decades (ranging
from 7.2 to 200 years), and are usually well-constrained with the currently available set of
photometric data. Thanks to the automatic surveys OGLE and MACHO, the time span is typically more
than 20 years. The eccentricities are rather mild (mostly below 0.1, median of about 0.06).

We also detected altogether six systems showing besides the classical apsidal motion also some
additional variation of their orbital periods. These six systems are shown in Fig. \ref{FigOClite}
together with their light-time effect fits, while the parameters of these fits are given in Table
\ref{LITEparam}. One can see that the periods are adequately short, being still well-covered with
the data.
 All of the systems are definitely stable ones, according to the stability criteria as published
earlier (see e.g. \citealt{2001MNRAS.321..398M}, or \citealt{2002A&A...384.1030S}). In publications
like \cite{2004RMxAC..21....7T} there were presented several different empirical stability criteria
with slightly different coefficients. However, all of our systems are located well bellow all of
these stability limits in the $P_3/P$ versus $e_3$ diagram, hence our sample cannot be used for
distinguishing between them.
 However, the more interesting dynamical effects of the third-body
mechanics \citep{2016MNRAS.455.4136B} are generally very small (see the last columns in Table
\ref{LITEparam} and Section \ref{results} above). If we expect to find some dynamical influence of
the third bodies on the eclipsing pairs (i.e. changing the inclination of the binary), it should be
visible after several more decades or even a century of observations for two most promising systems
(OGLE-SMC-ECL-1634 and OGLE-SMC-ECL-2152). For all of these systems a significant fraction of the
third light was also detected during the LC analysis (see Table \ref{LCOCparam}), which is an
indirect evidence that our hypothesis is credible.
 We are aware of the fact that the whole analysis is based on assumption that the mass ratio of the
 eclipsing system is equal to 1.0. However, as we have tested for a few systems, the main results
 would be shifted when assuming different mass ratio, but the result about non/detection of the
 third light remains.
Two other systems (namely OGLE-SMC-ECL-1985, and OGLE-SMC-ECL-4923) possibly also show additional
third-body variations, but the orbital periods are probably longer ($>$20 yr), hence we cannot do
any reliable analysis yet.

The number of such potential third-body systems is increasing every year, even outside of our
Galaxy, but our detected systems cannot be taken seriously as some comparative benchmark for
statistics. This is mainly due to the fact that in our sample of stars we simply preferred these
systems with more interesting variations in the $O-C$~diagrams. Hence the number of such potential
triples is apparently higher, than should be in real stellar population. However, the same apply
also for the other apsidal motion systems in our sample, which were preferably selected only when
having adequately short apsidal periods. In total, we have initially checked 35 obviously eccentric
SMC systems, among which 21 were selected for the present publication due to their adequately short
apsidal motion periods ($U < 200$~yr).

And finally, quite remarkable situation arises when we compare the number of known system in and
out of our own Galaxy. \cite{2018ApJS..235...41K} in their database lists in total 139 eccentric
systems where the precession of omega angle is detectable within our Galaxy. However, thanks to
recent papers on eccentric binaries in Magellanic Clouds (such as \cite{2014A&A...572A..71Z},
\cite{2015AJ....150..183Z}, \cite{2015AJ....150....1H}, \cite{2016MNRAS.460..650H}, or the present
study) we can surely say that nowadays we know more such systems outside of our own Galaxy!

\begin{table}[t]
\begin{center}
\caption{Relativistic apsidal motion and observed internal structure constant, log $k_{2,obs}$ for
seven selected systems.  } \label{tabK2}
\begin{tabular}{llll}
\hline\hline\noalign{\smallskip}
System & $\dot{\omega}_{rel}$ & $\dot{\omega}_{rel}/\dot{\omega}$ & log $k_{2,obs}$  \\
       & [deg/cycle]  &            [\%]                           &                  \\
\noalign{\smallskip}\hline\noalign{\smallskip}
 OGLE-SMC-ECL-1214 &  0.00186 &  6.15 &  -1.855  \\
 OGLE-SMC-ECL-1370 &  0.00243 &  4.23 &  -1.932  \\
 OGLE-SMC-ECL-1393 &  0.00204 &  2.80 &  -2.021  \\
 OGLE-SMC-ECL-1634 &  0.00221 &  3.90 &  -1.970  \\
 OGLE-SMC-ECL-1874 &  0.00235 &  2.80 &  -2.034  \\
 OGLE-SMC-ECL-2152 &  0.00154 &  3.25 &  -2.131  \\
 OGLE-SMC-ECL-2194 &  0.00299 &  1.68 &  -1.974  \\
\noalign{\smallskip}\hline
\end{tabular}
\end{center}
\end{table}

\begin{acknowledgements}
We are grateful to the anonymous referee for his/her fruitful comments and remarks, which greatly
improved the whole manuscript. This work was supported by the grant MSMT INGO II LG15010. We also
do thank the OGLE and MACHO teams for making all of the observations easily public available. We
are also grateful to the ESO team at the La Silla Observatory for their help in maintaining and
operating the Danish telescope. This research has made use of the SIMBAD and VIZIER databases,
operated at CDS, Strasbourg, France and of NASA Astrophysics Data System Bibliographic Services.
\end{acknowledgements}

%
%


\end{document}